# Exciton-coherence generation through diabatic and adiabatic dynamics of Floquet state

Kento Uchida[1,*], Satoshi Kusaba[1], Kohei Nagai[1], Tatsuhiko N. Ikeda[2], and Koichiro Tanaka[1,3,*]

[1]*Department of Physics, Graduate School of Science, Kyoto University, Sakyo-ku, Kyoto 606-8502, Japan*

[2] *Institute for Solid State Physics, University of Tokyo, Kashiwa, Chiba 277-8581, Japan*

[3] *Institute for Integrated Cell-Material Sciences, Kyoto University, Sakyo-ku, Kyoto 606-8501, Japan*

*\*e-mail: uchida.kento.4z@kyoto-u.ac.jp, kochan@scphys.kyoto-u.ac.jp*

## Abstract

Floquet engineering of electronic systems is a promising way of controlling quantum material properties on an ultrafast time scale. So far, the energy structure of Floquet states in solids has been observed through time and angle-resolved photoelectron spectroscopy or pump-probe measurement techniques. However, the dynamical aspects of the photon-dressed states under ultrashort pulse driving have not been explored yet. Their dynamics become highly sensitive to the envelope of the driving field when the light-matter interaction enters non-perturbative regime, and thus, understanding of them is crucial for ultrafast manipulation of quantum state. Here, we observed coherent exciton emissions under intense and non-resonant mid-infrared fields in monolayer $WSe_2$ at room temperature, which is unexpected in perturbative nonlinear optics. Together with numerical calculations, our measurements revealed that the coherent exciton emission reflects the diabatic and adiabatic dynamics of Floquet states. Our results provide a new approach to probing the dynamics of the Floquet state and lead to control of quantum materials through pulse shaping of the driving field.

**Main part**

Intense light waves are a useful tool for transforming functionalities of electronic systems on an ultrafast timescale through nonlinear light-matter interactions [1-4]. When the light wave is periodic in time (i.e., an ideal continuous wave), the driven electronic system is well described by the Floquet states [5]. They are "photon-dressed states" reflecting the effective Hamiltonian of the driven system, and their physical properties can be tuned by parameters such as the amplitude, frequency, and polarization of the light wave. Such a tuning capability has given rise to the concept of Floquet engineering, one of the most promising candidates for ultrafast manipulation of quantum systems [6-9]. The Floquet states under periodic driving $F(t) = F_0 \cos \Omega t$ are characterized by a ladder-like energy structure given by $\varepsilon^*(F_0, \Omega) + n\hbar\Omega$ ( $\varepsilon^*$: quasienergy, $n$: integer, $\Omega$: driver angular frequency) according to the Floquet theorem (Fig. 1(a)). Such a quasienergy structure has been observed using several ultrafast experimental techniques in solids [10-12].

However, the above Floquet picture assumes a continuous-wave driving ($F_0 = const.$) and hence is approximate in real experiments, where ultrashort pulses are usually used to yield intense fields. Instead, for finite-width pulses, the Floquet states arise only instantaneously at the time-dependent amplitude ($F_0 = F_0(t)$) [13-15]. These instantaneous Floquet states and their quasienergies gradually change along the pulse envelope $F_0(t)$, and the quantum state of the system either adiabatically follows or diabatically jumps between them. For the ultimate Floquet engineering, it is crucial to control those dynamics of the Floquet states during the pulse. Nevertheless, direct or indirect signatures of them have not yet been experimentally obtained in solids because of the fast dephasing of electrons in solids and the necessity of a huge quasienergy shift comparable to driving photon energy for observable effects. The lack of an established method that can access coherent electron dynamics during the pulse duration also makes it challenging to observe the dynamics of Floquet states.

Electronic systems driven by intense light waves emit coherent electromagnetic radiation, which entails the dynamical information of the system, and can hence be indirect evidence of instantaneous Floquet states. A typical example is high-order harmonic generation (HHG), in which radiation is emitted in integer multiples of the driver photon energy [16-21]. HHG can be regarded as an emission process from the Floquet state of electrons [7,22], reflecting the sub-cycle (short-time) dynamics imprinted in the ladder-like quasienergy structure separated by $\hbar\Omega$. On the other hand, to address the adiabatic and diabatic dynamics among instantaneous Floquet states, one needs to access the effect of long-time dynamics during the pulse on nonlinear emission, which cannot be described by conventional HHG process.

Here, we propose that nonlinear emissions from a driven system can be utilized as a probe of the dynamics of Floquet states. Let us consider the simplest model embodying this concept: a two-level system excited by an intense laser pulse showing transitions between two instantaneous Floquet states. Figure 1(a) shows the quasienergy structure of the two-level system as a function of the field amplitude $F_0$. We assume that the envelope of the driving field $F_0(t)$ varies slowly enough that the Floquet states are useful instantaneous basis: the electronic system at $t_0$ can be described by the Floquet state with lightwave amplitude $F_0(t = t_0)$. The validity of this assumption was experimentally confirmed for short pulses [23]. When the driving field amplitude is sufficiently weak, the transition probability as a function of incident photon energy has a local maximum at the well-known multi-photon resonance condition: $\Delta\varepsilon = n\hbar\Omega$, where $\Delta\varepsilon$ is the transition energy. When the field amplitude is strong, however, the energy levels shift to be quasienergies, and the multi-photon resonance condition reads

$$\Delta\varepsilon^*(\Omega, F_0) \approx n\hbar\Omega, \tag{1}$$

where $\Delta\varepsilon^*$ is the quasienergy difference between two Floquet states. This condition means that the two quasienergies approach to show an avoided crossing as illustrated in the inset of Fig. 1(a), where the diabatic transition between the two Floquet states efficiently occurs.

Since the quasienergy depends on the amplitude of the driving field, the resulting transition probability between two Floquet states intricately depends on the time profile of the field amplitude $F_0(t)$ [13-15]. As $F_0$ increases, the initial ground state $|g\rangle$ adiabatically transforms into the Floquet state $|g^*\rangle$, approaching the first avoided crossing with the Floquet state $|e^*\rangle$ (STEP 1 in Figs. 1). For brevity, we call $|g^*\rangle$ and $|e^*\rangle$ the Floquet ground and excited states, respectively. When the peak driving field amplitude is below the critical amplitude $F_c$ satisfying Eq. (1), the system does not reach the first avoided crossing and, thus, returns to the original ground state after the pulse duration without making a transition. On the other hand, when the peak electric field strength exceeds the critical field $F_c$, a diabatic transition from the Floquet ground to an excited states occurs with transition probability $P$, as predicted by solving the Floquet-Landau-Zener problem [13-15] (STEP 2 in Figs. 1). This transition creates a superposition (coherence) between two Floquet states oscillating at $\Delta\varepsilon^*/\hbar + n\Omega$. After the transition, as the field strength $F_0$ gradually decreases, the created coherence adiabatically evolves into the bare state coherence (STEP 3 in Fig. 1). This means that high harmonic emissions oscillating at integer multiples of the driver frequency are gradually converted into a coherent emission resonant with the transition of the original system as a signature of the diabatic and adiabatic evolution of the Floquet state. In perturbative nonlinear optics [24], coherent emissions are not allowed except at integer multiples of the driver frequency. The non-perturbative light-matter interaction, which alters the energy structure of the driven system, plays a crucial role in the above phenomenon. Moreover, the transition probability depending on $F_0(t)$ provides us with a knob for controlling the Floquet states.

To observe the dynamics of Floquet states in solids, we focused on excitons in semiconducting transition metal dichalcogenides (TMDCs). An exciton is a bound state of an electron-hole pair formed by Coulomb attraction. Due to it being a bound state, its dephasing time is much longer than that of the unbound electron-hole pairs. In particular, monolayer TMDCs are direct gap semiconductors and host excitons with a large binding energy of several hundred meV, which are stable even at room temperature [25-29]. Under strong infrared light driving, excitons in monolayer TMDCs form a Floquet state and show large energy shifts called the optical Stark shift and Bloch-Siegert shift [30-32]. In addition, efficient HHG and the related phenomena has been observed in a group of monolayer TMDCs [33-37]. Thus, stable excitons and their strong coupling to light provide us an ideal platform to observe the dynamical aspects of the Floquet states.

We prepared a monolayer WSe$_2$ sample by using the mechanical exfoliation technique and attached it to a quartz substrate (see the Supplemental Information). Figure 2(a) shows a schematic diagram of the experimental setup. We used an intense linearly polarized mid-infrared (MIR) driver to excite the sample. The MIR photon energy of 0.26 eV is far below the A-exciton resonance at room temperature (1.65 eV), which was confirmed by making a photoluminescence measurement (Fig. 2(b)). The A-exciton resonance is also far from any integer multiples of the driving photon energies (harmonic emission energies), which is suitable for observing the conversion of high harmonics into bare excitonic coherence. The pulse width was estimated to be 60 fs (~six-cycle pulse) by using the sum-frequency generation technique [36]. In accordance with previous experimental and theoretical investigations [23,15], we assumed that Floquet state are useful instantaneous basis in this experimental condition. The maximum intensity at the focal point was estimated to be 170 GW cm$^{-2}$ in air, corresponding to an electric field strength of 11 MV cm$^{-1}$. Under such a strong electric field, the A-exciton tranforms into a Floquet state, and its resonance (quasienergy) is expected to be shifted higher, on the order of 0.1 eV for the peak intensity [30-32,38]. This huge quasienergy shift enables the seven-photon resonance condition to be satisfied for dressed A-excitons ($n = 7$ for Eq. (1)). The emissions induced by the driver were collected by using a UV silica lens with an aperture to resolve the divergence of the emission beam. The polarization of the collected emissions was resolved by using a wire grid polarizer, and emission spectra were then detected by using a spectrometer equipped with a CCD camera. All the experiments were performed at room temperature in ambient conditions.

Figure 2(c) shows typical emission spectra from monolayer WSe$_2$ for an MIR electric field in the armchair direction. We observed clear high harmonics signals at $n\hbar\Omega$. The polarization of the high harmonics was parallel to the incident MIR polarization, as determined from the crystal symmetry [33,34]. In addition, we observed an emission at the A-exciton resonance with parallel and perpendicular components. One possible origin of the resonant emission is incoherent emission from excitons. So far, there have been several reports on luminescence from incoherent excitons created by intense terahertz or MIR pulses associated with high harmonics [16,17]. However, in our experiment, the intensity of the parallel component was larger than that of the perpendicular one. Since monolayer $WSe_2$ obeys the circular polarization selection rule, the information on the linearly polarized driving field is usually lost through the decoherence process. Thus, the partially polarized exciton emission strongly suggests that a non-negligible emission may result from excitonic coherence created by the dynamics of Floquet states.

To confirm the existence of excitonic coherence, we have measured temporal and spatial coherence of the exciton emission. From the polarization selection rule mentioned above, we can detect the temporal coherence of the exciton by the degree of polarization of the exciton emission. In addition, it is expected that the coherent exciton emission should have a small beam divergence that is observed in the spatially coherent signals such as the observed high harmonics, while the incoherent exciton emission has a large beam divergence regardless of the space-time coherence of the MIR driver. Figure 3(a) shows the emission spectra with open- and closed- aperture configurations. In Fig. 3(b) is shown a schematic setup for polarization-resolved exciton emission measurements with aperture. We collected emissions with a numerical aperture of 0.56 (0.06) for the open (closed) configuration. When the emission has coherence reflecting the wavefront of the driving field, the beam divergence becomes small compared with that from an incoherent emission such as luminescence. For example, since high harmonic generation is a coherent optical process, the divergence of the HHG signals is small enough to detect all the signals even in the close-aperture configuration. On the other hand, in the exciton emission, where there is a contribution from incoherent emission, the intensity decreases when the aperture is closed.

Here, we define the degree of linear polarization (DOLP) as $(I_{para} - I_{perp})/(I_{para} + I_{perp})$, where $I_{para}$ and $I_{perp}$ are the exciton emission intensities of the parallel and perpendicular components, respectively. By closing the aperture, the DOLP of the exciton emission should increase. Figure 3(c) shows the polarization states of exciton emissions in open-aperture (orange squares) and closed-aperture (blue circles) configurations, respectively. By closing the aperture, the DOLP changes from 0.4 to 0.7, indicating the space-time coherence of the exciton. The above hypothesis is supported by the results of a similar experiment we conducted on thin-layer bulk $WSe_2$. Since the luminescence efficiency is strongly suppressed in the bulk $WSe_2$ due to the indirect bandgap, we expected only coherent emissions from bulk $WSe_2$. We found that the DOLP of the exciton emission was almost 1.0 in bulk $WSe_2$ (see the Supplemental Information for a detailed discussion).

Note that we ruled out any valley coherence contribution to the exciton emission which can be also attributed to a partially linear polarized emission and has been reported in monolayer TMDCs mainly at low temperature [39]. Figure 3(d) shows the polarization state of luminescence measured by a pico-second laser excitation at 1.97 eV (630 nm) at room temperature. We could not find any features of valley coherence within the signal-to-noise ratio, indicating that the emissions with high DOLP originated only from the excitonic coherence at room temperature. Therefore, we can distinguish the contribution of the excitonic coherence by subtracting the perpendicular components from the parallel ones.

Figures 4(a) and (b) show the distinguished coherent and incoherent emission spectra near the excitonic resonance. Both emissions grow with MIR intensity. The saturation of the coherent excitonic emission at higher intensities (> 140 GW $cm^{-2}$) suggests that the driven system reaches the first avoided crossing between Floquet states, described by Eq. (1). Compared with the incoherent emission spectrum, the coherent emission peak is slightly blue-shifted and has a spectral component between the exciton resonance and the seventh harmonic. In addition, we can see spectral distortion in the seventh harmonic. The seventh harmonic peak shows a redshift just below the intensities where strong coherent exciton emissions emerge. Such a complex spectral change depending on MIR intensity was not seen in the fifth harmonic (see the Supplementary

Information) and suggests the effect of the Floquet state dynamics depicted in Figs. 1.

We simulated the exciton response under non-resonant driving by numerically solving a simple two-level model considering an electron-hole vacuum and the 1s level of the A-exciton in monolayer $WSe_2$ (see the Supplemental Information for a detailed description). Figure 4(c) shows the calculated emission spectra near the excitonic resonance with several field amplitudes near the critical field $F_c$. The simulation reproduced the spectral features of the observed coherent exciton emissions, including its slight blue shift. In addition, the simulated seventh order harmonics show a spectral modification similar to that of the experiments, i.e., an energy shift depending on driver intensity. This indicates that the observed excitonic response can be regarded as that of the simple two-level system depicted in Figs. 1: intense MIR driving induces a huge A-exciton resonance shift of ~ 0.15 eV at maximum intensity, creating excitonic coherence as a signature of the dynamics of the Floquet state. Note that we also solved the linearized semiconductor Bloch equation for the two-band model, which takes the excitonic effect into account and obtained qualitatively the same results (see the Supplemental Information) [40].

To extract the dynamics of the nonlinear emissions, we performed a Gabor transformation on the calculated emission (interband polarization), as shown in Fig. 4(d). The Gabor transformation enables us to track the temporal evolution of the energy of the nonlinear emissions with finite temporal and energy resolutions. The fifth harmonic signal (the blue line), whose emission energy is below the exciton resonance, shows an almost instantaneous response to the envelope of the driving field $F_0(t)$ (grey area in the upper panel), which is consistent with the conventional HHG response below the bandgap energy (HHG originating from Kerr-type nonlinearity) [41].

On the other hand, the seventh harmonic signal (the light-green line in Fig. 4(d)) is delayed relative to the envelope of the driving field $F_0(t)$. Since the electronic polarization is delayed when its frequency is nearly resonant with a certain transition (see the Supplemental Information for an explanation), the delay in the seventh harmonic indicates its resonance with the dressed A-excitons. At the bare excitonic resonance energy, the amplitude of the emission grows as the driving field amplitude decreases (the orange line in Fig. 4(d)). This evolution reflects the energy shift of nonlinear emission from that of seventh harmonic to the bare excitonic resonance. This energy shift clearly follows the time evolution of the quasienergy of the dressed exciton (the black dashed line), which means that the Floquet state transformed adiabatically into the bare exciton state under our experimental conditions (see the Supplemental Information for additional information). The adiabatic dynamics of the Floquet state leads to the characteristic of coherent exciton emission in Figs. 4(a) and (c): slight blue shift of spectral peak, spectral component between the seventh harmonic and the excitonic resonance.

The simulations revealed that the pulse width of the driver and the coherence time of the exciton are crucial factors to observing the coherent dynamics of the Floquet state. We need an excitonic coherence time longer than the falling edge of the driver pulse. The intensity ratio of the excitonic emission to the seventh harmonic is exponentially suppressed as the damping constant (inverse of the excitonic coherence time) increases (see the Supplemental Information for a detailed discussion). By using the spectral width of the A-exciton resonance in a sample from the same batch that was encapsulated with h-BN flakes, which strongly suppress inhomogeneous broadening, we estimated the homogenous broadening due to phonon scattering to be 25 meV at room temperature in the sample of this experiment [42,43]. This value is relatively consistent with the simulation, which reproduces the experimental results. This much longer coherence time of the exciton compared with unbound electron-hole pairs enables us to observe a clear signature of the dressed exciton dynamics in solids even at room temperature.

In summary, we demonstrated that coherent exciton emissions are induced by intense MIR driving in monolayer $WSe_2$ at room temperature as a signature of the Floquet state dynamics. Coherent exciton emissions far away from the harmonic emission energies indicate a dynamical energy structure change exceeding 0.1 eV, a diabatic transition between the Floquet states of the vacuum and the 1s A-exciton, and an adiabatic transformation of the Floquet states into bare A-exciton states. Our study provides a new method to access the Floquet state and indicates that monolayer TMDC is a fascinating platform for realizing coherent control of quantum properties on an ultrafast time scale. Time- and energy-resolved measurements of the coherent emission using ultrashort pulses will enable direct observation of Floquet state dynamics, as depicted in Fig. 4(d). For a peak driving field amplitude higher than in our experimental conditions, multiple avoided crossings between Floquet states are expected to be involved in the temporal evolution of the system. In this situation, the transition probability highly depends on the shape of the driving field pulse, owing to Floquet-Landau-Zener interference [15]. Therefore, pulse shaping of an intense driving field with control of light helicity might provide a knob for ultrafast control of the valley degree of freedom in monolayer TMDCs [35,44].

## Data availability

The data that support the findings of this study are available from the corresponding author upon reasonable request.

## Acknowledgements

This work was supported by Grants-in-Aid for Scientific Research (S) (Grant Nos. JP17H06124 and JP21H05017), JST ACCEL Grant (No. JPMJMI17F2). K. U. is thankful for a Grant-in-Aid for Young Scientists (Grant No. 19K14632).

S. K. was supported by MEXT Quantum Leap Flagship Program (No. JPMXS0118067634). K. N. was supported by a JSPS fellowship (Grant No. JP20J14428). T. N. I. was supported by JSPS KAKENHI Grant No. JP21K13852.

## Author contributions

K. U. conceived the project. K. T. supervised the project. S. K. prepared the sample, and evaluated the characteristic of the sample by luminescence measurement. K. U. and K. N. constructed the setup for the nonlinear emission measurements. K. U. carried out the nonlinear emission measurement and data analysis. K. U. carried out the numerical calculation with the help of K. N.. All authors contribute to the discussion and the writing of the manuscript.

**Additional information**

**Supplementary Information** accompanies this paper.

**Competing financial interests**

The authors declare no competing financial interests.

**Figure captions**

**Figure 1. Temporal evolution of Floquet state in two-level system.** **(a)** Schematic diagram of Floquet energy bands as a function of field strength $F_0$. Here, the solution of time-dependent Schrödinger equation is defined by $|i^*\rangle = \exp(-i\varepsilon_i^* t/\hbar) \sum |i^*, l\rangle \exp(il\Omega t)$, where $\varepsilon_i^*$ is quasienergy, $|i^*, l\rangle$ is the $l$th sideband state of Floquet state $|i^*\rangle$. Blue (red) lines represent ladder-like energy levels of the ground (excited) Floquet state separated by a driver photon energy of $\hbar\Omega$. The depth of color represents the magnitude squared of the Floquet sidebands given by $\langle i^*, l|i^*, l\rangle$ with logarithmic scale. The inset shows an expanded view of the avoided crossing between the Floquet ground and excited states. Here, the label of Floquet states ($i^* = g^*$ or $e^*$) is assigned in diabatic picture. **(b)** Schematic evolution of quantum state under slowly varying field

amplitude $F_0(t)$. **STEP 1**. The initial ground state $|g\rangle$ transforms into a Floquet state $|g^*\rangle$ and shows an energy shift with increasing $F_0$. **STEP 2**. At the first avoided crossing, a diabatic transition from $|g^*\rangle$ to $|e^*\rangle$ occurs, as shown in the inset. **STEP 3**. The created superposition between the ground and excited Floquet states $(\alpha|g^*\rangle + \beta|e^*\rangle)$ oscillating at $7\Omega$ adiabatically transforms into a superposition of bare state $(\alpha|g\rangle + \beta|e\rangle)$ oscillating at $\Delta\varepsilon/\hbar$, where $\Delta\varepsilon=\varepsilon_e - \varepsilon_g$.

**Figure 2. Observation of nonlinear emission process in monolayer WSe$_2$ at room temperature.** **(a)** Experimental setup of nonlinear emission measurement. The spot size of the MIR pulse at the focal point is estimated to be 30 μm (full width at half maximum). All the experiments were performed at room temperature in an ambient condition. **(b)** Typical nonlinear emission spectra from the monolayer WSe$_2$ sample. Light blue (orange) indicates the spectrum component whose polarization is parallel (perpendicular) to the MIR polarization. Here, the MIR polarization direction is along the armchair direction of monolayer WSe$_2$, which is determined by the selection rule of high harmonic generation [33,34]. MIR intensity is estimated to be 170 GW cm$^{-2}$.

**Figure 3. Characteristics of excitonic resonant emission.** **(a)** Nonlinear emission spectra near the A-exciton resonance with open (green) and closed (light blue) aperture configurations. **(b)** Schematic exciton emission from the sample. **(c)** Polarization-resolved intensities of the excitonic resonant emission with open (green squares) and closed (blue circles) aperture configurations. Here, 0 degree is the direction parallel to the MIR polarization. **(d)** Polarization-resolved intensities of photo luminescence from A-exciton created by 1.97 eV continuous wave excitation. Here, 0 degree is the direction parallel to the polarization of the excitation laser.

**Figure 4. Spectral evolution of the excitonic resonant emission depending on MIR intensity.** **(a, b)** Measured nonlinear emission spectra of **(a)** coherent and **(b)** incoherent components near the A-exciton resonance with several MIR

intensities. The MIR polarization is along the armchair direction. **(c)** Computed nonlinear emission spectra near the resonance in the two-level model. The detailed conditions of the calculation are described in the Supplemental Information. **(d)** Upper panel: Temporal profile of the driving field amplitude. Lower panel: Emission intensity of two-level model as a function of time and harmonic order obtained by using Gabor transformation. Here, the time window of the Gabor transformation is 20 fs. Blue, orange, and green solid lines shows the emission intensities at the fifth harmonic, the excitonic resonance, and the seventh harmonic, respectively. The maximum intensities are normalized for clarity. The black dashed line represents the calculated quasienergy of dressed the A-exciton in the two-level model. Here, the quasienergy at time $t$ is evaluated by using the field amplitude of $F_0(t)$.

Fig. 1

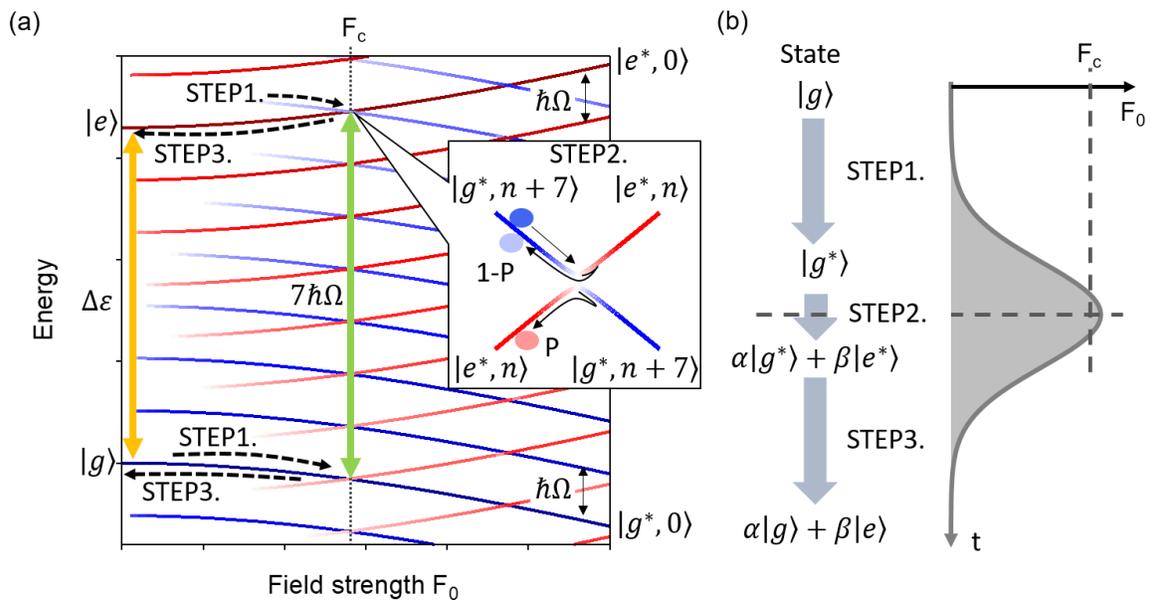

Fig.2

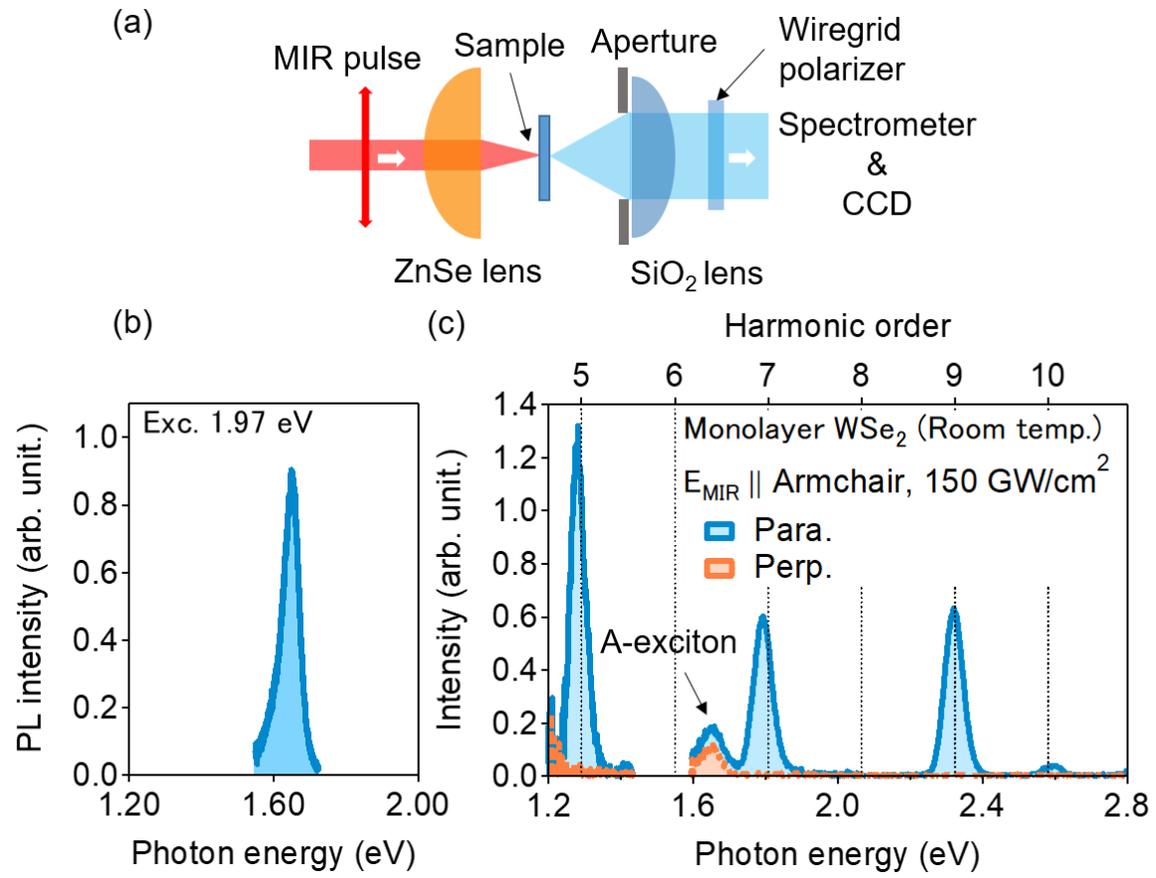

Fig.3

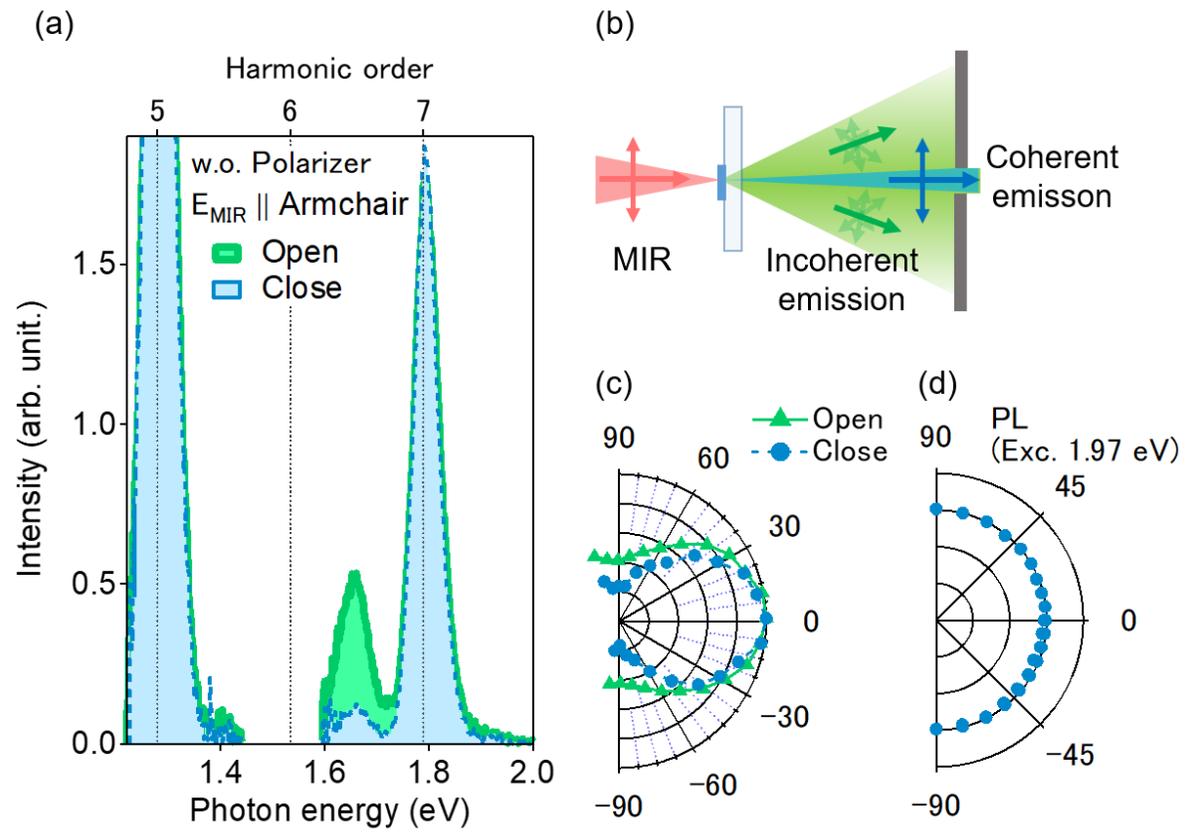

Fig.4

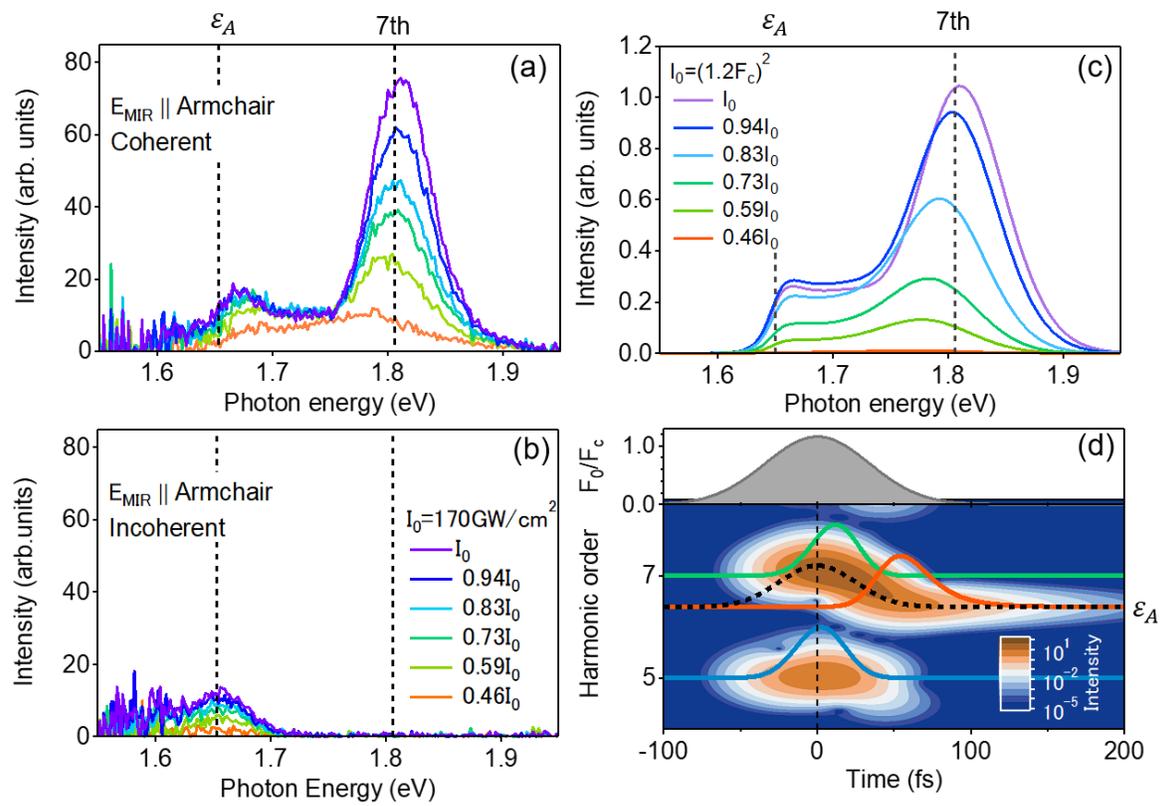

# Supplemental Information:

# Exciton-coherence generation through diabatic and adiabatic dynamics of Floquet state

Kento Uchida[1,*], Satoshi Kusaba[1], Kohei Nagai[1], Tatsuhiko N. Ikeda[2], and Koichiro Tanaka[1,3,*]

[1]Department of Physics, Graduate School of Science, Kyoto University, Sakyo-ku, Kyoto 606-8502, Japan

[2] Institute for Solid State Physics, University of Tokyo, Kashiwa, Chiba 277-8581, Japan

[3] Institute for Integrated Cell-Material Sciences, Kyoto University, Sakyo-ku, Kyoto 606-8501, Japan

*e-mail: uchida.kento.4z@kyoto-u.ac.jp, kochan@scphys.kyoto-u.ac.jp

## 1. Sample preparation.

We prepared monolayer $WSe_2$ by mechanical exfoliation of a commercial $WSe_2$ crystal grown by chemical vapor deposition (CVD), which was purchased from 2D semiconductors Inc. The monolayer was transferred onto the quartz substrate by dry transfer method. The sample was characterized by photoluminescence (PL) spectrum obtained by a commercial PL/Raman micro spectrometer (NanoFinder 30, Tokyo Instruments Inc.).

Figure S1 shows the optical image of the exfoliated WSe$_2$ sample. The size of the monolayer region is ~200 μm, which is sufficiently large for the MIR laser spot size of ~60 μm (white dashed circle).

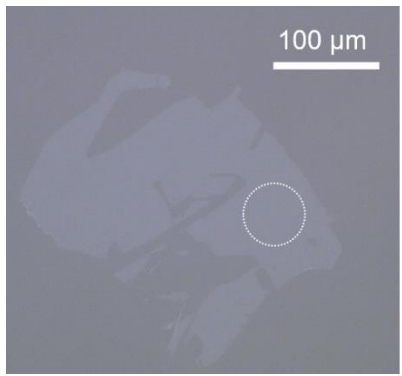

Figure S1: Optical image of the mechanically exfoliated monolayer WSe$_2$ sample. White dashed circle indicates the spot where MIR light was irradiated. The sample size is sufficiently larger than the MIR spot size, enabling us to obtain high signal-to-noise ratio signals.

2. **Polarization-resolved PL measurement setup.**

The polarization-resolved PL measurement was performed in the backscattering configuration. We used a commercial supercontinuum white light source (WhiteLase, Fianium, total power: 4 W, repetition: 40 MHz, wavelength: 400~2400 nm). A home-built subtractive-mode double monochromator and a shortpass filter was used to obtain the excitation light with narrow linewidth. The excitation light was focused onto the sample by using an objective lens (Mitsutoyo, 20x, N.A.=0.28). The PL was collected by the same objective lens, polarization-resolved by using a film polarizer, spectrally resolved by a grating spectrometer, and detected by a liquid-nitrogen-cooled Si CCD camera.

3. **Nonlinear emission measurement setup.**

The laser source is a Ti: Sapphire regenerative amplifier (pulse width: 35 fs, pulse energy: 7 mJ, center wavelength: 800nm, repetition rate: 1 kHz). Part of laser output was used for optical parametric amplification to generate signal and idler outputs in near-infrared region using Light Conversion TOPAS-C. MIR pulses (wavelength: 4800 nm) for nonlinear emission measurement were generated by using difference frequency mixing of signal and idler outputs. The signal and idler outputs were then blocked by a longpass filter (cutoff wavelength: 4000 nm). The MIR polarization angle and intensity were controlled by two wire-grid polarizers (Thorlabs WP25M-IRA) and a liquid crystal variable retarder (Thorlabs LCC1113-MIR). The MIR pulses were focused onto the sample with normal incidence by using an ZnSe lens (focal length: 62.5 mm). The group delay dispersion due to the above optics was compensated by inserting a pair of $CaF_2$ plate. The spot size at the focal point was estimated to be 60 μm in full-width at half-maxima (FWHM) by using knife edge measurement, and the pulse width is estimated to be 60 fs by using the sum frequency mixing of the fundamental and the MIR pulses in monolayer $MoS_2$ [S1]. The transmitted nonlinear emissions were corrected by UV-fused silica lens (focal length: 50 mm). To evaluate the beam divergence of the nonlinear emission from the sample, we set an iris in front of the collection lens so that we can control the numerical aperture for beam collection. The collected emissions were polarization-resolved by using a wire-grid polarizer (Thorlabs WP25M-UB), spectrally resolved by a spectrometer, and detected by a Si CCD camera. The crystal orientation of the sample was determined by the selection rule of HHG in monolayer $WSe_2$ [S2].

## 4. Nonlinear emission measurement in bulk $WSe_2$.

Figure S2(a) shows typical emission spectra from bulk 2H-WSe$_2$ corresponding to that from monolayer in Fig. 2(c). Since 2H-WSe$_2$ has an inversion symmetry, there are no even-order harmonics observed in the monolayer sample. Similar to the monolayer sample, we observed emissions slightly below the seventh order harmonic (~ 1.6 eV), which is probably attributed to the resonant emission from direct excitons of bulk WSe$_2$. In bulk WSe$_2$, fast energy relaxation of direct exciton into indirect exciton disturbs the luminescence process. Therefore, we can expect the only coherent exciton emission from the bulk sample. In fact, the perpendicular component of the nonlinear emission is negligible. Figure S2(b) shows the polarization state of the exciton emission. The emission is polarized parallel to incident MIR polarization, and its DOLP reaches 0.97. These results in bulk WSe$_2$ support our claim in the main text: excitonic coherence originated from Floquet state dynamics is involved in the exciton emission under non-resonant MIR field.

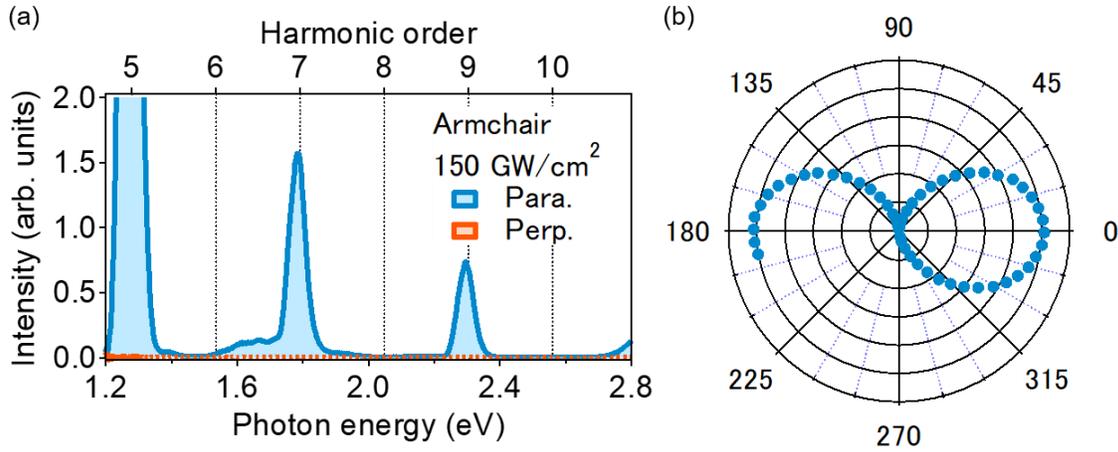

Figure S2: (a) Typical nonlinear emission spectra from the bulk WSe$_2$ sample. Light blue (orange) indicates the spectrum component whose polarization is parallel (perpendicular) to the MIR polarization. (b) Polarization-resolved intensities of the exciton emission. Here, 0 degree is the direction parallel to the MIR polarization.

## 5. MIR intensity dependence of the fifth harmonic

Figure S3(a) shows the fifth harmonic spectra for several MIR intensities. Compared with the seventh harmonics shown in Fig. S3(b), spectral peak energy of the fifth harmonic is almost independent of MIR intensity. This suggests that the strong spectral modification of the seventh harmonic is caused by the Floquet state dynamics of A-exciton in monolayer $WSe_2$.

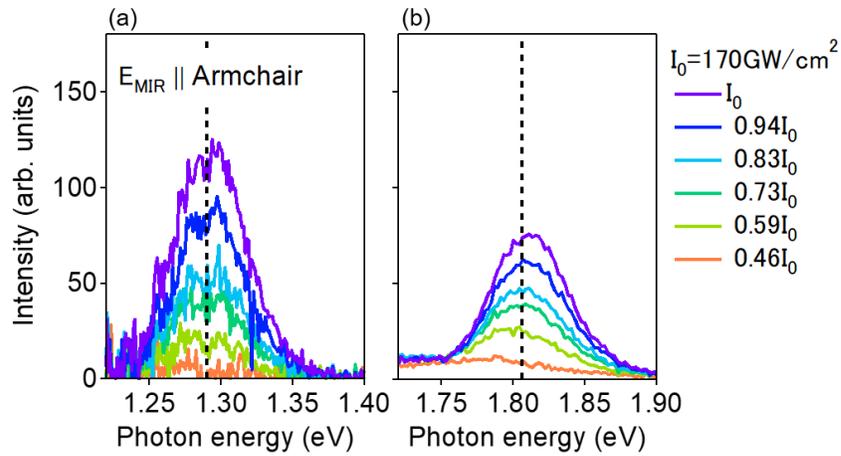

Figure S3: (a) The fifth harmonic spectra with several MIR intensities. (b) The seventh harmonic spectra with several MIR intensities.

## 6. Numerical simulation of the excitonic response under intense MIR driving.

Here, we consider the excitonic response under the intense MIR field by numerically solving (i) the simplified two-level model and (ii) two-band model with the excitonic effect.

### 6-1. Numerical simulation of the dynamics of two-level model.

Figure S4(a) shows the energy diagram of the two-level system. In this subsection, for simplicity, we consider only the vacuum state for electron-hole pairs $|0\rangle$ and the 1s state of A-exciton $|1s\rangle$, and neglects higher excitonic states and electron-hole continuum band. Then, Hamiltonian of the driven system can be described as follows:

$$H(t) = H_0 + H_I(t), \tag{S1}$$

$$H_0 = \begin{pmatrix} 0 & 0 \\ 0 & \varepsilon_{1s} \end{pmatrix}, \tag{S2}$$

$$H_I(t) = \begin{pmatrix} 0 & d_{1s}F(t) \\ d_{1s}F(t) & 0 \end{pmatrix}. \tag{S3}$$

Here, $\varepsilon_{1s}$ is the transition energy of 1s excitonic energy (1.65 eV), $d_{1s}$ is the transition dipole moment of 1s exciton, and $F(t)$ is the temporal profile of the driving field. Here, the light-matter coupling is considered within the dipole-approximation, and $d_{1s}$ is a real number for simplicity. Then, the temporal evolution of the density matrix obeys the so-called optical Bloch equations:

$$\hbar \frac{\partial}{\partial t} \rho_{0,0}(t) = 2\hbar \Omega_R(t) Im[\rho^*_{1s,0}(t)], \tag{S4}$$

$$\rho_{1s,1s}(t) = 1 - \rho_{0,0}(t), \tag{S5}$$

$$\hbar \frac{\partial}{\partial t} \rho_{1s,0}(t) = (-i\varepsilon_{1s} - \hbar\gamma)\rho_{1s,0}(t) - i\hbar\Omega_R(t)(\rho_{0,0}(t) - \rho_{1s,1s}(t)), \tag{S6}$$

$$\rho_{0,1s}(t) = \rho^*_{1s,0}(t), \tag{S7}$$

Here, $\Omega_R(t) = d_{1s}F(t)$, $\rho_{ij}$ is the $(i,j)$ component of the density matrix, and $\gamma$ is a phenomenological damping constant. The induced polarization $P(t)$ is given by

$$P(t) = 2d_{1s}Re[\rho_{1s,0}(t)]. \tag{S8}$$

The nonlinear emission spectrum is given by $\omega^2|\tilde{P}(\omega)|^2$, where $\tilde{P}(\omega)$ is the Fourier transformation of $P(t)$. In the numerical simulation, we used Gaussian pulse given by

$$F(t) = F_0 \exp(-t^2/\Delta t^2) \cos \Omega t. \tag{S9}$$

We set the parameters as $\varepsilon_{1s} = 1.650$ eV, $\hbar\Omega = 0.258$ eV, $\hbar\gamma = 0.020$ eV, $\Delta t = 36$ fs. Here, the Rabi energy is given by $\hbar\Omega_R = d_{1s}F_0$, and the initial conditions are $\rho_{0,0} = 1$ and $\rho_{1s,1s} = \rho_{1s,0} = 0$. Figure S4(b) shows a typical nonlinear emission spectrum calculated using the two-level model with $\hbar\Omega_R = 0.6$ eV, corresponding to $0.12\,F_c$ in the main text. In addition to the high harmonic generation, strong exciton emission is observed. Figure S4(c) shows the temporal evolution of the polarization obtained by the Gabor transformation.

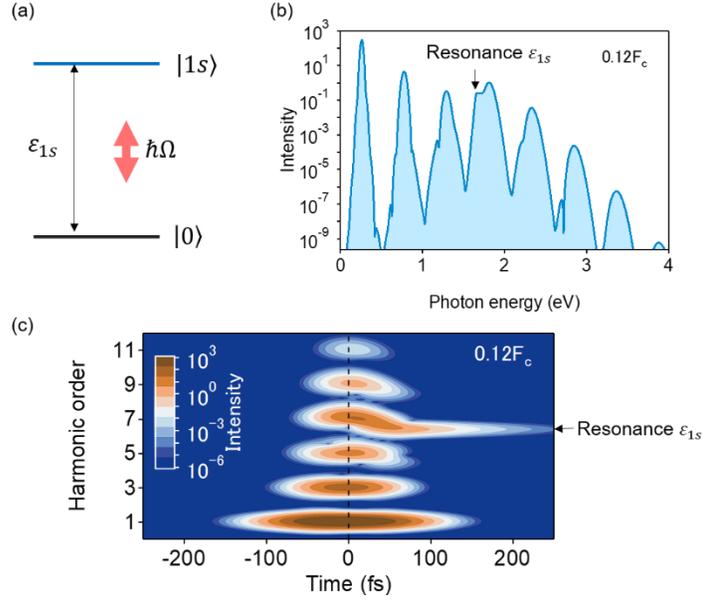

Figure S4: (a) The energy diagram of two-level model consisting of vacuum and 1s exciton levels. (b) The calculated nonlinear emission spectra by using two level model. The arrow indicates the 1s exciton transition energy. (c) Temporal evolution of nonlinear emission obtained by the Gabor transformation with the time window of 20 fs (FWHM).

After the peak of driving pulse ($t > 0$), the seventh harmonic signal is gradually converted into the bare excitonic emission.

### 6-2. Floquet states in a two-level model.

When the driving field is a continuous wave $F(t) = F_0 \cos\Omega t$, we can apply the Floquet theorem [S3], and the solution of the time-dependent Schrödinger equation can be written by

$$\Phi_\alpha(t) = e^{-\frac{i\varepsilon_\alpha^*}{\hbar}t} \sum_l \phi_\alpha^{(l)} e^{-il\Omega t}, \tag{S10}$$

$$\phi_\alpha^{(l)} = a_{\alpha,0}^{(l)} \varphi_0 + a_{\alpha,1s}^{(l)} \varphi_{1s}, \tag{S11}$$

where $\varepsilon_\alpha^*$ is quasienergy, $\phi_\alpha^{(l)}$ is $l$th sidebands of the Floquet state $\Phi_\alpha(t)$. $\phi_\alpha^{(l)}$ can be described as a superposition of bare states $\varphi_\beta$ with a coefficient $a_{\alpha,\beta}^{(l)}$.

The set of a quasienergy $\varepsilon_\alpha^*$ and the coefficients $a_{\alpha,\beta}^{(l)}$ can be obtained by solving the static problem as follows [4]:

$$\sum_m (H^{(l-m)} - m\Omega \delta_{lm}) \phi_\alpha^{(m)} = \varepsilon_\alpha^* \phi_\alpha^{(l)}, \tag{S12}$$

$$\sum_l H^{(l)} e^{-il\Omega t} = H_0 + \frac{\Omega_R}{2}(e^{i\Omega t} + e^{-i\Omega t})\begin{pmatrix} 0 & 1 \\ 1 & 0 \end{pmatrix}, \tag{S13}$$

In the calculation, we numerically solved Eq. (S12) by considering the Fourier component of $-100 \leq m, l \leq 100$. Figure S5(a) shows the numerically calculated quasienergy diagram as a function of the Rabi energy (field amplitude). With the increase of Rabi energy, quasienergy is monotonically shifted as shown in Fig. S5(b). Here, we neglect the avoided crossing of quasienergy, and trace the energy with the diabatic picture. This energy shift considerably deviates from that of the perturbation-theory calculation including ac-Stark shift and Bloch-Siegert shift typically above $\hbar\Omega_R = 0.2$ eV. The intersection between Floquet sidebands, which actually shows an avoided crossing, represents the multi-photon resonance conditions in the Floquet system given by Eq. (1) in the main text. In our experimental setup, the first avoided crossing at Rabi energy around 0.6 eV satisfies the seven-photon resonance condition. This energy corresponds to the field amplitude where excitonic coherence is efficiently created as shown in Figs. S4(b) and (c).

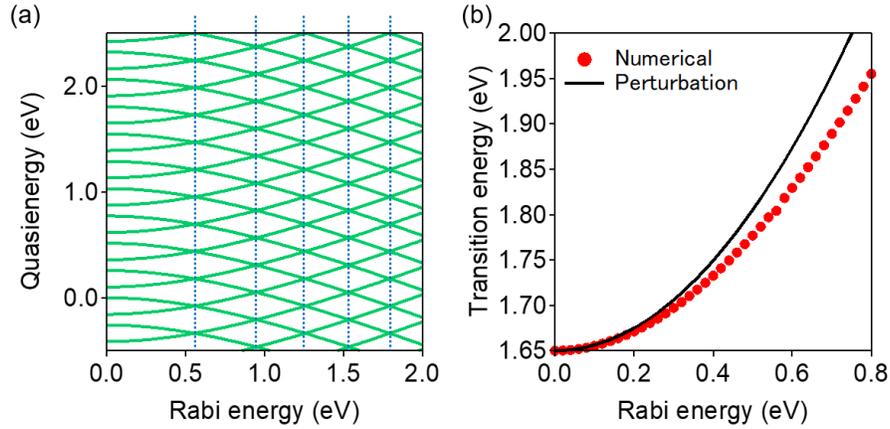

Figure S5: (a) Numerically obtained quasienergy diagram of the two level model. Here, $\varepsilon_{1s} = 1.65$ eV and $\hbar\Omega = 0.258$ eV. The dashed lines indicate Rabi energies corresponding to the avoided crossing. (b) Transition energy shifts between two levels as a function of Rabi energy. Red circles indicate the numerically obtained one by neglecting avoided crossing, and black solid line indicates analytical one by considering up to the second-order perturbation of the interaction Hamiltonian. This energy shift within the perturbation theory corresponds to so-called ac-Stark shift and Bloch-Siegert shift.

### 6-3. Temporal evolution of Floquet state

Let us consider the dynamics of the system under temporally varying envelope $F_0(t)$ using instantaneous Floquet basis. Here, we assume that the quantum state at time $t_0$ can be described by the Floquet state under the constant driving amplitude $F_0(t_0)$ (instantaneous Floquet state) [S4,S5], and the transition between different instantaneous Floquet states only occurs at the avoided crossing of quasienergies. The system is in the ground state (vacuum state) at the initial time. Then, with an increase of the field amplitude, the ground state becomes the Floquet ground state as follows:

$$\Psi(t) = \Phi_{g*}(t) = e^{-i\theta_{g*}(t)} \sum_l \phi_{g*}^{(l)}(F_0(t)) e^{-il\Omega t}, \qquad (S14)$$

$$\theta_{g*}(t) = \int_0^t dt' \frac{\varepsilon_g^*(F_0(t'))}{\hbar}. \qquad (S15)$$

Then, the induced electric polarization is given by

$$P(t) = -\langle \Psi(t) | e\hat{x} | \Psi(t) \rangle$$
$$= d_{1s} \sum_{l,m} \left[ a_{g*,0}^{(l)*}(F_0(t)) a_{g*,1s}^{(m)}(F_0(t)) e^{i(l-m)\Omega t} + c.c. \right]$$
$$= 2 d_{1s} \sum_{n,m} A_{g*,n,m}(F_0(t)) \cos(n\Omega t + \psi_{g*,n,m}(F_0(t))), \qquad (S16)$$

$$A_{g*,n,m}(F_0(t)) = \left| a_{g*,0}^{(n+m)*}(F_0(t)) a_{g*,1s}^{(m)}(F_0(t)) \right|, \tag{S17}$$

$$\psi_{g*,n,m}(F_0(t)) = \arg\left[ a_{g*,0}^{(n+m)*}(F_0(t)) a_{g*,1s}^{(m)}(F_0(t)) \right]. \tag{S18}$$

Equation (S16) indicates that the induced electric polarization has integer multiple frequency components with slowly varying amplitude and phase, i.e., high harmonic emissions.

When the peak field amplitude is below the first avoided crossing, the quantum state returns to the ground state, and only high harmonic emission with slight modulation can be observed. On the other hand, when the multi-photon resonance condition of the Floquet state is satisfied, the diabatic transition between ground and excited Floquet states efficiently occurs [S5,S6].

Thus, the quantum state after the transition becomes a superposition of Floquet ground and excited states as follows:

$$\Psi(t) = c_{g*} \Phi_{g*}(t) + c_{e*} \Phi_{e*}(t). \tag{S19}$$

Then, the induced electric polarization is given by

$$P(t) = 2d_{1s} \sum_{\alpha,n,m} |c_\alpha|^2 A_{\alpha,n,m} \cos(n\Omega t + \psi_{\alpha,n,m})$$
$$+ d_{1s} \left[ c_{g*}^* c_{e*} e^{i\Delta\theta} \sum_{n,m} \left( a_{g*,0}^{(n+m)*} a_{e*,1s}^{(m)} + a_{g*,1s}^{(n+m)*} a_{e*,0}^{(m)} \right) e^{in\Omega t} + c.c. \right], \tag{S20}$$

where $\Delta\varepsilon^*(F_0(t))$ and $\Delta\theta(F_0(t))$ are the transition energy and phase difference between the two Floquet states, respectively. The first term describes high harmonic emissions as previously discussed. On the other hand, the second term has an extra fast-oscillating phase $\Delta\theta(F_0(t))$, which corresponds to the frequency of $\Delta\varepsilon^*/\hbar$ at time t. Especially, when the field amplitude decreases, absolute values of $a_{g*,0}^{(0)}$ and $a_{e*,1s}^{(0)}$ becomes much larger than those of other coefficients $a_{\alpha,\beta}^{(l)}$. Therefore, at each instant of time t, the second term is approximately written by

$$d_{1s}\left[c_{g*}^* c_{e*}\left(a_{g*,0}^{(0)*} a_{e*,1s}^{(0)}\right)e^{-\frac{i\Delta\varepsilon^*(F_0(t))}{\hbar}t+const.} + c.c.\right]. \tag{S21}$$

This means that excitonic coherence with the transition energy of Floquet state $\Delta\varepsilon^*(F_0(t))$ is created after the transition. Finally, after the pulse duration, $a_{\alpha,\beta}^{(l)}$ becomes zero except for $a_{g*,0}^{(0)} = a_{e*,1s}^{(0)} = 1$. Then the induced electric polarization after pulse duration is given by

$$P(t) = d_{1s}\text{Re}\left[c_{g*}^* c_{e*} e^{\frac{i\varepsilon_{1s}}{\hbar}t+const.}\right]. \tag{S22}$$

This means that the emission with the 1s exciton resonance energy can be observed after the pulse duration.

According to the above description, let us consider our experimental condition, i.e., the transition energy of the Floquet state at the peak field amplitude is almost resonant with the seventh order harmonic energy. Before the peak of the driving pulse, high harmonic emission can be observed as in Eq. (S16). Then, around the peak of the driving pulse, the dressed excitonic coherence oscillating with the seventh order harmonic energy ($\Delta\varepsilon^* = 7\hbar\Omega$) is created. With a decrease of driving field amplitude, dressed excitonic coherence energy gradually approaches the bare 1s exciton energy according to $\Delta\varepsilon^*(F_0(t))$. Figure S6 shows the time evolutions of transition energy $\Delta\varepsilon^*(F_0(t))$ (red dashed line) and peak energy of the numerically calculated emission in the vicinity of the 1s excitonic energy. As expected, the emission energy is almost the same as the seventh order harmonic energy before the peak, and follows $\Delta\varepsilon^*(F_0(t))$ after the peak. This indicates that our experimental condition is well described by the instantaneous Floquet states.

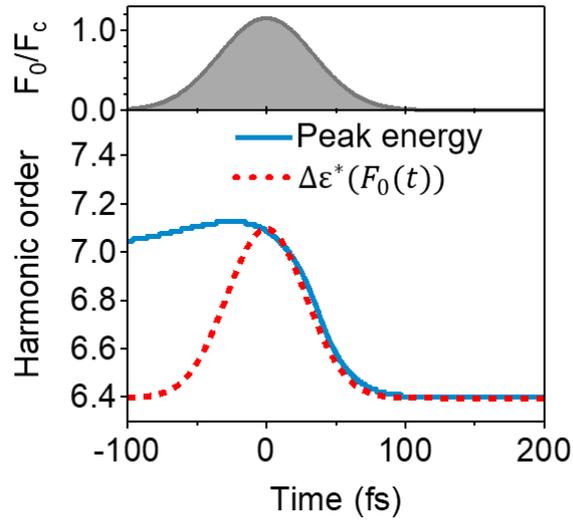

Figure S6: Upper panel: temporal profile of the driving field amplitude. Lower panel: Temporal evolutions of transition energy and emission energy in the vicinity of 1s excitonic resonance $\varepsilon_{1s}$ (~6.4 in the figure). Blue solid curve indicates the peak energy of emission extracted from the Gabor transformation of polarization $P(t)$ in Fig. S4(c). The slight deviation from the seventh harmonic emission energy before the peak of the pulse is maybe due to the phase $\psi_{g*}$ in Eq. (S16). Red dashed curve indicates the transition energy of Floquet state $\Delta\varepsilon^*(F_0(t))$. According to the time window of Gabor transformation of 20 fs, the transition energy is also averaged over 20 fs in the figure for clarity.

## 6-3. Multi-photon resonance condition in the Floquet state.

To check our claim that coherence between two Floquet states is efficiently created at multi-photon resonance $\Delta\varepsilon^*(F_0, \Omega) \approx n\hbar\Omega$, we performed numerical simulations of the two-level system with the different peak amplitudes (Rabi

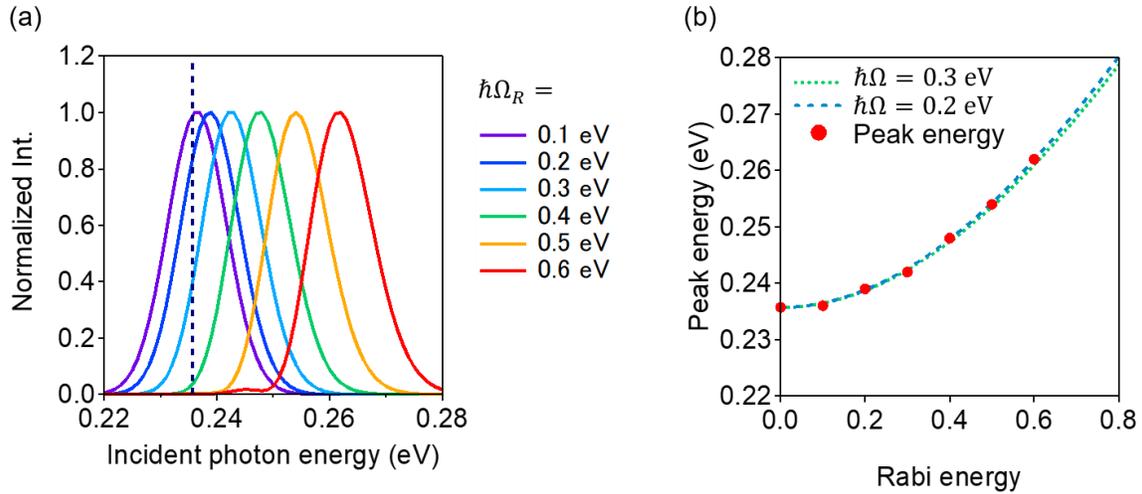

Figure S7 (a) Emission intensities at 1.65 eV (1s excitonic resonance) as a function of driving photon energy. The intensities are normalized by the local maximum value near the seven-photon resonance. Black dashed line indicates seven-photon resonance of bare 1s exciton ($\hbar\Omega = \varepsilon_{1s}/7$). To separate the resonant emission component from the seventh order harmonic, we use Gabor window centered at 230 fs with 20 fs width, which is sufficiently after the pulse duration. (b) Red line: peak energy extracted from (a). Green dotted and blue dashed lines indicate the transition energies of the Floquet state divided by 7 ($\Delta\varepsilon^*/7$) with $\hbar\Omega = 0.3$ eV and 0.2 eV, energy) and driving photon energies. Figure S8(a) shows driving photon energy dependence of the resultant emission intensity resonant with the bare transition energy. When $\hbar\Omega_R$ is equal to 0.1 eV, the intensity becomes maximum almost at the seven-photon resonance of the bare 1s excitonic energy $\hbar\Omega = \varepsilon_{1s}/7$ (dashed line in Fig. S7(a)). With an increase of the peak field amplitude (Rabi energy), the peak energy is shifted toward higher energy. At $\hbar\Omega_R = 0.6$ eV, the peak energy is almost 30 meV higher than the original seven-photon resonance. Figure S7(b) shows the driving photon energy where the resonant emission takes the maximum value as a function of Rabi energy. The peak energy (red circles) follows the transition energies of the Floquet state (green dotted line: $\hbar\Omega = 0.3$ eV, blue dashed line: $\hbar\Omega = 0.2$ eV) well. This clearly indicates the multi-photon resonance condition for Floquet state described in Eq. (1) in the main text.

## 6-4. Effect of decoherence on the exciton emission.

Here, we verify the effect of decoherence on the exciton emission. Figure S8(a) shows the exciton emission and the seventh harmonic intensity as a function of the dephasing rate $\hbar\gamma$. With an increase of the dephasing rate, both the seventh harmonic and the exciton emission rapidly decrease. The decreasing rate of the exciton emission is larger than that of the seventh harmonic. Figure S8(b) shows the intensity ratio of the exciton emission to the seventh harmonic. To observe the exciton emission comparable to the seventh harmonic, the dephasing rate should be as small as possible. The dephasing rate of 1s A-exciton in $WSe_2$ is estimated to be around 20 meV even at room temperature. This small dephasing rate enables us to observe exciton coherence created through Floquet state formation in our experimental condition.

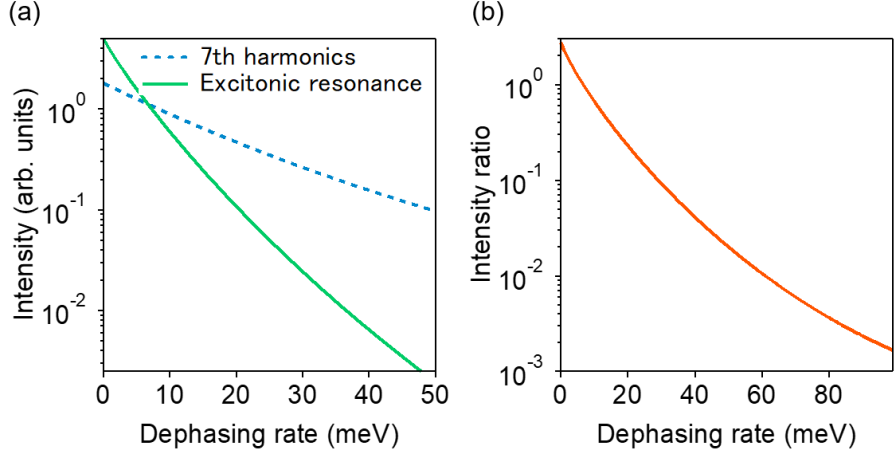

Figure S8 (a) Emission intensities at 1.65 eV (1s excitonic resonance: green solid line) and 1.81 eV (seventh harmonic: blue dashed line) as a function of dephasing rate $\hbar\gamma$. The intensities are taken from the peak value of Gabor transformation of polarization at each photon energy. (b) Intensity ratio of the excitonic emission to the seventh harmonic as a function of dephasing rate.

### 6-5. Two-band model with excitonic effect.

Here, we consider the two-band model with excitonic effect in two-dimensional momentum space. The treatment is basically the same as Refs. [S7] and [S8]. We numerically solved the linearized semiconductor Bloch equations as follows [S9]:

$$\frac{d}{dt}\tilde{P}_{\boldsymbol{k}}(t) = -\gamma\tilde{P}_{\boldsymbol{k}}(t) - \frac{i}{\hbar}\varepsilon_g(\boldsymbol{k} - e\boldsymbol{A}(t)/\hbar)\tilde{P}_{\boldsymbol{k}}(t) - i\Omega_{\mathrm{R},\boldsymbol{k}}(t)(2\tilde{n}_{\boldsymbol{k}}(t) - 1) + \sum_{\boldsymbol{q}\neq\boldsymbol{k}}V_{\boldsymbol{k}-\boldsymbol{q}}\tilde{P}_{\boldsymbol{q}}(t)\,, \tag{S23}$$

$$\frac{d}{dt}\tilde{n}_{\boldsymbol{k}}(t) = -\frac{2}{\hbar}\mathrm{Im}[\Omega_{\mathrm{R},\boldsymbol{k}}(t)\tilde{P}_{\boldsymbol{k}}^*(t)] \tag{S24}$$

$$\boldsymbol{A}(t) = -\int dt\,\boldsymbol{F}(t) \tag{S25}$$

Here, $\tilde{P}_{\boldsymbol{k}}(t)$ and $\tilde{n}_{\boldsymbol{k}}(t)$ are the pair function and population at crystal momentum $\boldsymbol{k}$, respectively. $\varepsilon_g(\boldsymbol{k})$ and $\Omega_{\mathrm{R},\boldsymbol{k}}(t)$ are crystal-momentum dependent bandgap energy and Rabi energy, respectively. For simplicity, we used two-dimensional cosine bands $\varepsilon_g(\mathbf{k}) = \varepsilon_g(\mathbf{0}) + t_r(1 - \cos k_x a \cos k_y a)$ ($\varepsilon_g(\mathbf{0}) = 1.89$ eV: bandgap energy, $t_r = 3.5$ eV: hopping energy, $a = 3.3 \times 10^{-10}$ m: lattice constant), and neglected momentum dependence of Rabi energy ($\Omega_{R,\mathbf{k}}(t) = \Omega_R(t)$). $V_{\mathbf{k}-\mathbf{q}}$ is the two-dimensional Coulomb potential between electrons at $\mathbf{k}$ and $\mathbf{q}$. We set the amplitude of $V_{\mathbf{k}-\mathbf{q}}$ so that 1s exciton energy obtained by solving the Wannier equation is the same as the experimentally obtained one (1.65 eV). Here, $\Omega_{R,\mathbf{k}}(t)$ describes the creation and annihilation of electron-hole pairs, and $\varepsilon_g(\mathbf{k} - e\mathbf{A}(t)/\hbar)$ describes intraband motion of electron-hole pairs. To check the effect of intraband electron-hole motion on the excitonic response, we introduce a scaling parameter $\beta$ as $\varepsilon_g(\mathbf{k} - e\beta\mathbf{A}(t)/\hbar)$ in Eq. (S23). $\beta$ determines the strength of intraband electron-hole motion respective to the strength of interband mixing described by Rabi energy. In the numerical simulation, we used a $100 \times 100$ $k$-mesh and obtained the absorption spectrum as shown in Fig. S9(a). Figure S9(b) shows the nonlinear emission spectra with several $\beta$ and Rabi energy of $\Omega_R \sim 0.6$ eV. When $\beta = 1$, the typical kinetic energy of electron-hole pairs is estimated to be around 1.5 eV, which is larger than the Rabi energy, and the exciton emission is strongly suppressed. This might be because strong intraband driving of electron-hole pairs causes field ionization of 1s exciton into unbound electron-hole pairs. In contrast, when $\beta = 0$ (without intraband driving), strong exciton emission is observed as in the two-level model. Our experimental observation, which is similar to that with $\beta \sim 0$, suggests that the kinetic energy of electron-hole pairs is smaller than the Rabi energy and the excitonic response is well described by a two-level system. The more refined studies, including the real electronic structure, momentum-dependent Rabi energy, and scattering process, are needed to understand why we can regard excitons in monolayer WSe$_2$ as a two-level system well.

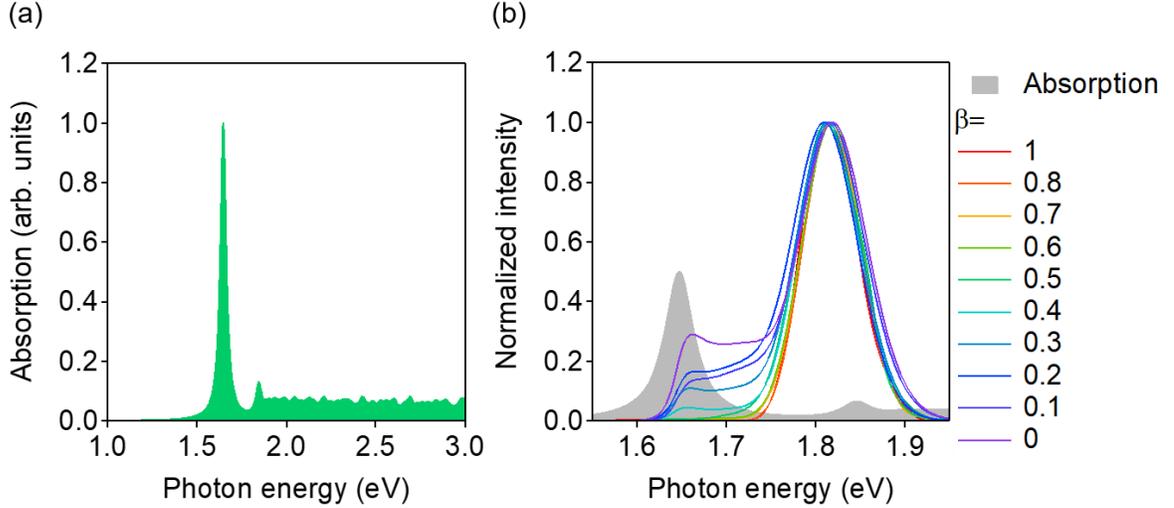

Figure S9 (a) Calculated absorption spectrum in two-dimensional cosine band with excitonic effects. Here, we solved Wannier equation with $100 \times 100$ k-mesh. (b) Nonlinear emission spectra near the excitonic resonance. Here, we set $\Omega_R = 0.65$ eV. Grey shaded area is the corresponding absorption spectrum.

## 7. Delayed response of electric polarization at resonance condition.

Here, we discuss the envelope of the electric polarization depending on incident frequency. For simplicity, we consider linear electric polarization described as follows:

$$P(t) = \int_{-\infty}^{t} dt' \, \chi(t-t') F(t'). \tag{S24}$$

As a response function, we consider the oscillator with exponential decay given by

$$\chi(t) = Ae^{-\gamma t} \sin \omega_0 t, \tag{S25}$$

where $A$, $\omega_0$, and $\gamma$ are the amplitude, resonant frequency, and damping constant of the oscillator, respectively. Here, we introduce the incident electric field given by

$$F(t) = F_0 \theta(t) \sin \omega t, \tag{S26}$$

$$\theta(t) = \begin{cases} 0 & (t < 0) \\ 1 & (t \geq 0) \end{cases}. \tag{S27}$$

Then, the resultant polarization is given by

$$P(t) = -\frac{AF_0}{2((\omega_0 - \omega)^2 + \gamma^2)} [\gamma(\cos \omega t - e^{-\gamma t} \cos \omega_0 t) - (\omega_0 - \omega) \times (\sin \omega t - e^{-\gamma t} \sin \omega_0 t)] + \frac{AF_0}{2((\omega_0 + \omega)^2 + \gamma^2)} [\gamma(\cos \omega t - e^{-\gamma t} \cos \omega_0 t) - (\omega_0 + \omega)(\sin \omega t + e^{-\gamma t} \sin \omega_0 t)] \quad (t \geq 0). \tag{S28}$$

First, let us consider the resonance condition $\omega = \omega_0$. In such case, the main contribution of the electric polarization is described by

$$P(t) \approx -\frac{AF_0}{2\gamma}(1 - e^{-\gamma t}) \cos \omega t. \tag{S29}$$

As shown in Fig. S10 (solid red line), the envelope of electric polarization at the resonance condition is gradually increase with a time-scale of $1/\gamma$. Next, let us consider the non-resonance condition $\gamma \ll |\omega - \omega_0|$. This results in the electric polarization given by

$$P(t) \approx \frac{AF_0}{\omega_0^2 - \omega^2}(\omega \sin \omega t - \omega_0 \sin \omega_0 t). \tag{S30}$$

In this case, the envelope of the signal oscillates with the detuning between the oscillator resonance and incident laser frequency $(\omega - \omega_0)$ (see solid blue line in Fig. S10). Therefore, the time scale of the build-up of electric polarization is given by $(\omega - \omega_0)^{-1}$.

This indicates that the response of the electric polarization becomes instantaneous with an increase of the detuning. This discussion can be applied to the nonlinear electric polarization, and the delayed seventh harmonic signal with respect to the envelope of the incident driving field as shown in Fig. 4(d) suggests that the seventh harmonic is resonant with the transition energy

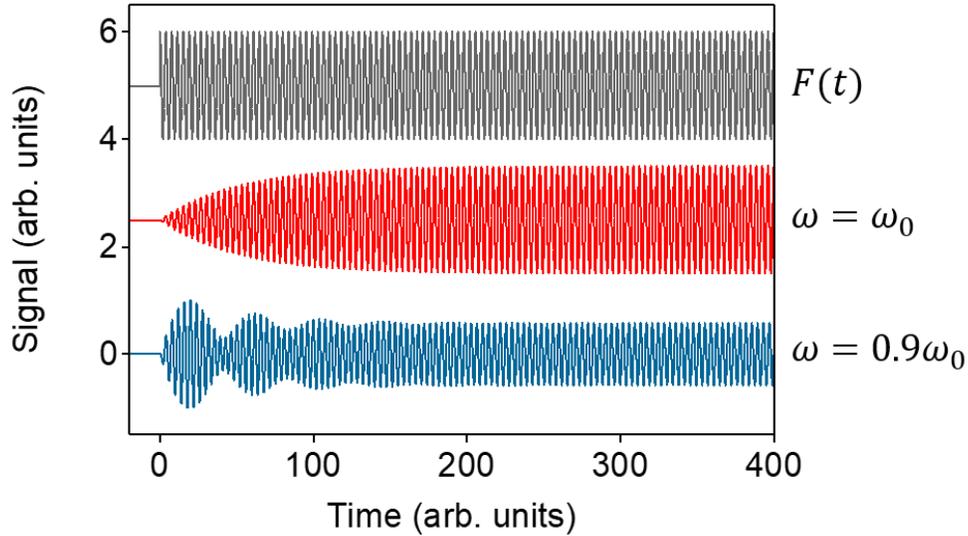

Figure S10 Electric polarization induced by the driving field with step-like envelope. Here, the parameters are set as follows: $\omega_0 = 1.65$ and $\gamma = 0.02$. Grey line indicates the temporal profile of the driving field $F(t)$. Red and blue solid lines indicate the electric polarization $P(t)$ with $\omega = \omega_0$ and $\omega = 0.9\omega_0$, respectively.

between Floquet states.